\documentclass[a4paper,twocolumn,english,prl, aps, showpacs]{revtex4}
\usepackage[T1]{fontenc}
\usepackage[latin9]{inputenc}
\usepackage{amsmath}
\usepackage{graphicx}
\usepackage{amssymb}

\makeatletter

\usepackage{babel}
\makeatother

\begin{document}

\title{Sequential Desynchronization in Networks of Spiking Neurons with
Partial Reset}

\author{Christoph Kirst$^{1,3}$, Theo Geisel$^{1-3}$ and Marc Timme$^{1,2}$}

\affiliation{$^{1}$Max Planck Institute for Dynamics and Self-Organization (MPIDS)
and\\
$^{2}$Bernstein Center for Computational Neuroscience (BCCN) G�ttingen,
37073 G�ttingen, Germany\\
$^{3}$Faculty of Physics, Georg-August-University G�ttingen, Germany}

\begin{abstract}
The response of a neuron to synaptic input strongly depends on whether
or not it has just emitted a spike. We propose a neuron model that
after spike emission exhibits a partial response to residual input
charges and study its collective network dynamics analytically. We
uncover a novel desynchronization mechanism that causes a sequential
desynchronization transition: In globally coupled neurons an increase
in the strength of the partial response induces a sequence of bifurcations
from states with large clusters of synchronously firing neurons, through
states with smaller clusters to completely asynchronous spiking. We
briefly discuss key consequences of this mechanism for more general
networks of biophysical neurons.
\end{abstract}

\pacs{05.45.Xt, 87.10.+e, 87.19.lj, 87.19.lm, 89.75.-k, 89.20.-a}

\maketitle
The brain processes information in networks of neurons, which interact
by sending and receiving electrical pulses called action potentials
or spikes. The response of a neuron to incoming signals strongly depends
on whether or not it has just sent a spike itself. After the initiation
of a spike the membrane potential at the cell body (soma) is reset
towards some potential and the response to further synaptic input
is reduced due to the refractoriness of the neuron \cite{Bio}. The
dendritic part of the neuron where incoming signals are integrated,
is affected only indirectly by this reset due to intra-neuronal interactions
\cite{Segev,Bressloff,Rospars}.

Several multi-compartment models have been proposed, in which different
parts of a single neuron interact to characterize this effect \cite{Segev}.
For instance, in a two-compartment model \cite{Bressloff} of coupled
dendrite and soma, the membrane potential at the soma is reset after
spike emission while the dendritic dynamics is affected only by the
resistive coupling from the soma to the dendrite. This accounts for
the fact that in several kinds of neurons residual charge remains
on the dendrite (following the somatic reset), that is then transferred
to the soma\cite{Rospars,DSCoupling}. Thus the dynamics of the
individual neurons is modified which severely affects the collective
capabilities of networks of such neurons.

In this Letter we propose a simple neuron model which captures the
response to residual input charges following spike emission in form
of a \emph{partial reset} and at the same time allows an analytical
study of the collective network dynamics.  A fraction $c\in[0,1]$
of the residual supra-threshold input charge is kept by the neuron
after reset. For $c=0$ all additional input charge not needed to
trigger a spike is lost after spike emission, whereas for $c=1$ the
total input charge is conserved \cite{c0c1}. Although the regime
$0<c<1$ is likely to be the biologically more relevant, to our knowledge,
there are so far no systematic studies of the dynamics of \emph{networks}
of neurons with partial response. To reveal the basic mechanisms underlying
the collective dynamics of networks of such neurons we focus on networks
of globally and homogeneously coupled neurons. Despite their simplicity
these networks already exhibit a rich variety of dynamics that is
controlled by the partial reset. In particular we find and show analytically
that for a broad class of neurons there is a desynchronization transition
in the network dynamics determined by a sequence of bifurcations:
For small $c$ the fully synchronous state coexists with a variety
of cluster states (cf. \cite{Cluster}), with differently sized groups
of synchronously firing neurons. With increasing $c$, states with
clusters of size $a$ and larger become sequentially unstable at bifurcation
points $c_{\mathrm{cr}}^{(a)}$ satisfying $0\leq\ldots\le c_{\mathrm{cr}}^{(3)}\leq c_{\mathrm{cr}}^{(2)}\leq1$
such that for sufficiently large $c>c_{\mathrm{cr}}^{(2)}$, only
an asynchronous state is left. We investigate the main mechanism generating
this sequence of bifurcations analytically and give an intuitive explanation.
We also discuss key consequences of this novel desynchronization mechanisms
for biophysically more detailed systems. 

\begin{figure}[b]
\begin{centering}
\includegraphics{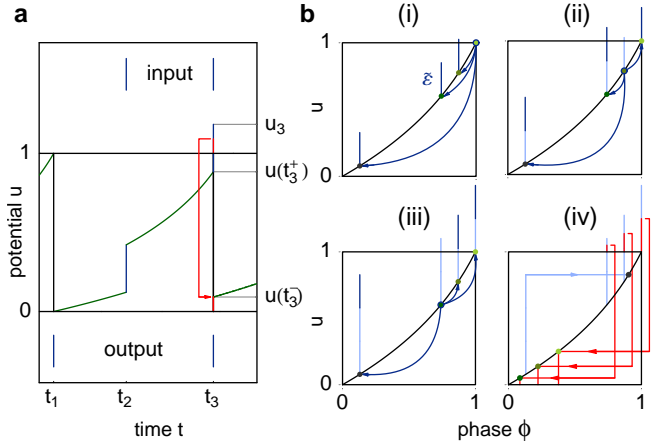}
\par\end{centering}

\caption{\label{fig: Model}Model Dynamics. \textbf{(a)} Membrane potential
$u$ of a single neuron reaches threshold by intrinsic dynamics at
time $t_{1}$, sends a spike and is reset. In response to a sub-threshold
input at $t_{2}$, the potential is increased instantaneously by $\tilde{\varepsilon}$.
At time $t_{3}$ the neuron receives supra-threshold input, $u\left(t_{3}^{-}\right)+\tilde{\varepsilon}\ge1$,
sends a spike and a residual fraction $c\in[0,1]$ of the charge,
not used to reach the threshold, remains and yields a partial reset
to the value $u\left(t_{3}^{+}\right)=c\left(u_{3}-1\right)$. \textbf{(b)}
Spike avalanche ($a=3$, $n=3$) in a network of $N=4$ neurons. \textbf{(i)}
The avalanche is triggered by neuron $1$ reaching the threshold (defining
the triggering set $\Theta_{0}=\{1\}$) and sending spikes to all
other neurons. \textbf{(ii)} This forces neuron $2$ to cross threshold
which then itself spikes ($\Theta_{1}=\{2\}$). \textbf{(iii)} In
turn, this induces a firing of neuron $3$ ($\Theta_{3}=\{3\}$),
completing this avalanche. \textbf{(iv)} Finally the new states of
all neurons are determined using the partial reset \eqref{eq: reset}
with $\Theta=\left\{ 1,2,3\right\} $.}

\end{figure}

Consider a network of $N$ oscillatory neurons \cite{MS}, whose
state at time $t$ is characterized by a phase variable $\phi_{i}$,
$i\in\{1,2,\dots,N\}$, that constantly increases with time $\frac{d}{dt}\phi_{i}=1.$
The membrane potentials $u_{i}=U\left(\phi_{i}\right)$ are specified
by a rise function $U$, that characterizes the subthreshold dynamics
of a neuron. Here $U$ is smooth, strictly monotonically increasing
($U'>0$) and normalized to $U(0)=0$ and $U(1)=1$. 

A neuron $j$ generates a spike when its membrane potential crosses
a threshold, $u_{j}(t^{-})\geq1$, which in turn may trigger an avalanche
of spikes (cf.~Fig.~\ref{fig: Model}): Neurons reaching the threshold
due to the free time evolution define the triggering set $\Theta_{0}=\left\{ j\:|\: u_{j}\left(t^{-}\right)=1\right\} $.
The units $j\in\Theta_{0}$ generate spikes which are instantaneously
received by all the neurons $i$ in the network. In response, their
potentials are updated according to \begin{equation}
u_{i}^{(1)}=u_{i}\left(t^{-}\right)+\sum_{j\in\Theta_{0}}\varepsilon_{ij}\label{eq: ui1}\end{equation}
where $\varepsilon_{ij}\geq0$ determines the strength of a directed
synaptic connection from neuron $j$ to $i$. The initial pulse may
trigger certain other neurons $k\in\Theta_{1}=\left\{ k\:|\: u_{k}\left(t^{-}\right)<1\le u_{k}^{(1)}\right\} $
to spike, etc. This process continues $n\le N$ steps until no new
neuron crosses the threshold. At each step $m\in\{2,3,\dots,n\}$
the potentials are updated according to \begin{equation}
u_{i}^{(m+1)}=u_{i}^{(m)}+\sum_{j\in\Theta_{m}}\varepsilon_{ij}\label{eq: uim}\end{equation}
where $\Theta_{m}=\left\{ k\:|\: u_{k}^{(m-1)}<1\le u_{k}^{(m)}\right\} $.
The phases immediately after the avalanche $\Theta=\bigcup_{q=0}^{n}\Theta_{q}$
of size $a=\left|\Theta\right|$ are obtained via\begin{equation}
\phi_{i}\left(t^{+}\right)=\begin{cases}
U^{-1}\left(u_{i}\left(t^{-}\right)+\sum_{j\in\Theta}\varepsilon_{ij}\right) & \, i\notin\Theta\\
U^{-1}\left(R\left(u_{i}\left(t^{-}\right)+\sum_{j\in\Theta}\varepsilon_{ij}-1\right)\right) & \, i\in\Theta\end{cases}\label{eq: reset}\end{equation}
where $R$ is the partial reset function. Here we focus on the linear
form $R\left(\zeta\right)=c\zeta$, with $c\in[0,1]$ specifying the
remaining fraction of supra-threshold input charges after reset.
 As a key example of the collective dynamics of neurons with partial
reset, we here study neurons with convex rise function ($U''>0$,
modelling e.g. a class of conductance based integrate-and-fire neurons),
which are homogeneously and globally coupled without self-interactions,
$\varepsilon_{ij}=\left(1-\delta_{ij}\right)\tilde{\varepsilon}$,
and total input strength $\varepsilon=\sum_{j}\varepsilon_{ij}=(N-1)\tilde{\varepsilon}<1$.

\begin{figure}[t]
\begin{centering}
\includegraphics{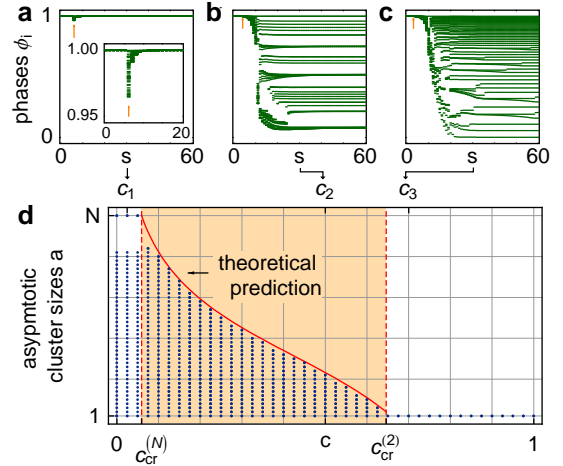}
\par\end{centering}

\caption{\label{fig:Convex}Sequential desynchronization transition in a network
of $N=50$ neurons ($U=U_{b}$, $b=-3$, $\tilde{\varepsilon}=0.0175$).
The phases $\phi_{i}$ of all neurons are plotted against the $s$-th
spike of a reference neuron. Starting from a synchronous state and
perturbing at $s=5$, the phase dynamics are shown for \textbf{(a)}
$c_{1}=0.025<c_{\mathrm{cr}}^{(N)}$ (inset: magnification), \textbf{(b)}
$c_{2}=0.5\in\left(c_{\mathrm{cr}}^{(N)},c_{\mathrm{cr}}^{(2)}\right)$
and \textbf{(c)} $c_{3}=0.7>c_{\mathrm{cr}}^{(2)}$. \textbf{(d)}
Observed cluster sizes (dots) in the asymptotic dynamics of $6000$
simulations for each $c$ value starting from different initial conditions.
red line: exact theoretical prediction \eqref{eq:cImplicit} above
which clusters are unstable. }

\end{figure}

Systematic numerical investigations indicate a strong dependence of
the network dynamics on the partial reset strength $c$: In particular,
we find synchronous states, cluster states, asynchronous states and
a sequential desynchronization of clusters with increasing $c$. More
detailed, if $c$ is sufficiently small, the long-term dynamics is
dominated by many coexisting cluster states in which neurons are synchronized
to differently sized groups resulting in a periodic state of the network
(cf. Fig. \ref{fig:Convex}). As $c$ increases, less and less clusters
are observed with the maximal number of units per cluster decreasing.
Above a critical value of $c$ only an asynchronous state remains.

What is the origin of this rich repertoire of dynamics and which mechanisms
control the observed transition? To answer these questions, we analytically
investigate the existence and stability of periodic states with clusters
of arbitrary sizes. It turns out that the sequence of bifurcations
is controlled by two effects: sub-threshold inputs that are always
synchronizing and supra-threshold inputs that may be synchronizing
or desynchronizing depending on the strength $c$ of the partial reset.

As the first step we show that the fully asynchronous (splay \cite{splay})
state exists and is stable for all $c\in[0,1]$. It is defined by
identical inter-spike intervals between each pair of subsequently
and individually firing neurons. A \emph{firing map} maps the phases
$\phi_{i}$ of the system just before one avalanche to the phases
just before the next. To construct this map for the asynchronous state
we evaluate the current spike (a 1-neuron {}``avalanche'') and shift
all phases by the same amount $\sigma$ such that the largest of the
resulting phases is at threshold. Without loss of generality, we label
the neurons' phases in ascending order such that the phases $\phi_{i}^{*}$
and the shift $\sigma^{*}$ uniquely define the asynchronous state;
they are determined self-consistently by $\phi_{1}^{*}=\sigma^{*}>0$
and $\phi_{l}^{*}=U^{-1}\left(U\left(\phi_{l-1}^{*}\right)+\tilde{\varepsilon}\right)+\sigma^{*}$
for $l\in\left\{ 2,\dots,N\right\} $ such that $\phi_{N}^{*}=1$.
Homogeneity of the network implies invariance of such an asynchronous
state under the firing map for every $\varepsilon<1$. Applying a
small perturbation $\boldsymbol{\delta}^{(0)}=\left(\delta_{1}^{(0)},\dots,\delta_{N-1}^{(0)}\right)$
to the $N-1$ phases which are not at threshold and linearizing the
firing map (cf. \cite{Kirst,Linear}) yields the perturbations after
the next firing \begin{equation}
\boldsymbol{\delta}^{(1)}=A\boldsymbol{\delta}^{(0)}.\label{eq:asynchA}\end{equation}
Here $A$ is a matrix whose only non-zero elements are $A_{i+1,i}=a_{i}$
for $i\in\left\{ 1,\dots N-2\right\} $ and $A_{i,N-1}=-a_{N-1}$
where \begin{equation}
a_{i}=\frac{U'\left(\phi_{i}^{*}\right)}{U'\left(U^{-1}\left(U\left(\phi_{i}^{*}\right)+\tilde{\varepsilon}\right)\right)}\label{eq:asynchAi}\end{equation}
for $i\in\left\{ 1,\dots N-1\right\} $. Since $U'>0$, $U''>0$,
and $\tilde{\varepsilon}>0$ we have $a_{i}<1$. The Enestr\"om -
Kakeya theorem \cite{Horn} applied to the matrix $A$ implies that
its eigenvalues satisfy $\left|\lambda_{i}\right|<1$ for all $i\leq N-1$,
showing that the asynchronous state is linearly stable. The stability
properties of this state are identical for all $c\in[0,1]$ because
all neurons fire individually and do not initiate any avalanche with
supra-threshold inputs.

Next we investigate the stability properties of a periodic cluster
state under the return map, i.e. the mapping of all phases just before
the triggering of an avalanche $\Theta$ to all phases just before
the same avalanche reoccurs. Such a cluster state exists (i.e., is
invariant) unless the maximal cluster size is too large such that
this cluster absorbs neurons not belonging to it or is absorbed by
other clusters. Given that the specific neuron $N\in\Theta_{0}$
stays in the triggering set of the avalanche, the return map $M$
equals the map defined form the hyperplane $\phi_{N}=1$ to itself.
It is fully specified by the number $m$ of avalanches, $1\le m\le N$,
by the number $a_{s}$, $s\in\left\{ 1,\dots,m\right\} $, of neurons
spiking in each avalanche, and by the subsequent phase shifts $\sigma_{s}$
that fix the time lags between the avalanches. This information is
determined from the initial phase vector $\left(\phi_{1},\dots,\phi_{N-1},1\right)$
and grouped into a firing sequence $\mathcal{F}=\left[\left(\varepsilon_{s},\sigma_{s}\right)\right]_{s=0}^{m}$,
setting $\varepsilon_{s}=a_{s}\tilde{\varepsilon}$. For given $\mathcal{F}$
the return map then reads \begin{equation}
M_{\mathcal{F}}\left(\phi_{i}\right)=S_{\sigma_{m}}\circ H_{\varepsilon_{m}}\circ\dots\circ S_{\sigma_{2}}\circ H_{\varepsilon_{2}}\circ S_{\sigma_{1}}\circ J_{\varepsilon_{1}}\left(\phi_{i}\right)\label{eq: return map}\end{equation}
for $i\in\Theta$. Here $S_{\sigma}(\phi)=\phi+\sigma$ is the map
mediating a pure phase shift, $H_{x}(\phi)=U^{-1}\left(U\left(\phi\right)+x\right)$
specifies the sub-threshold response to an incoming spike and $J_{x}(\phi)=U^{-1}\left(R\left(U\left(\phi\right)+x-\tilde{\varepsilon}-1\right)\right)$
represents the partial response $R$ to supra-threshold input. By
definition we have $M_{\mathcal{F}}(1)=1$. The conditions\begin{multline}
M_{\mathcal{F}}\left(\left[U^{-1}\left(1-a\tilde{\varepsilon}\right),U^{-1}\left(1-(a-1)\tilde{\varepsilon}\right)\right]\right)\\
\subset\left[U^{-1}\left(1-a\tilde{\varepsilon}\right),1\right]\label{eq: conditions}\end{multline}
for all $a\in\left\{ 1,\dots,a_{1}\right\} $ then ensure that all
neurons firing in the first avalanche $a_{1}$ will fire together
in an avalanche after return of neuron $N$ to threshold ($\phi_{N}=1$).
Thus \eqref{eq: conditions} ensure stability of a cluster $\Theta$
of size $|\Theta|=a_{1}$. For general $R$ and $U$ these conditions
yield upper and lower bounds \cite{Kirst} on the maximal size of
a cluster to be stable under the return map. Here we focus on the
specific rise function $U_{b}(\phi)=\frac{1}{b}\ln\left(1+\left(e^{b}-1\right)\phi\right)$,
$b<0$, where the change of phase differences due to sub-threshold
inputs is independent of the phase, i.e. $H_{\varepsilon}(\phi)-H_{\varepsilon}(\psi)=H_{\varepsilon}\circ S_{\sigma}(\phi)-H_{\varepsilon}\circ S_{\sigma}(\psi)$
for $\sigma\ge0$. For $i\in\Theta$ the return map \eqref{eq: return map}
then simplifies to  

\begin{equation}
M_{\mathcal{F}}\left(\phi_{i}\right)=S_{\bar{\sigma}}\circ H_{(N-a_{1})\tilde{\varepsilon}}\circ J_{a_{1}\tilde{\varepsilon}}\left(\phi_{i}\right)\label{eq: return U_b}\end{equation}
with $\bar{\sigma}=1-H_{(N-a_{1})\tilde{\varepsilon}}\circ J_{a_{1}\tilde{\varepsilon}}\left(1\right)$.
Since $M_{\mathcal{F}}'\ge0$ and $M_{\mathcal{F}}''\ge0$ the conditions
\eqref{eq: conditions} are all satisfied if the single condition
\begin{equation}
M_{\mathcal{F}}\left(U_{b}^{-1}\left(1-\tilde{\varepsilon}\right)\right)\ge U_{b}^{-1}\left(1-\tilde{\varepsilon}\right),\label{eq:MFUb}\end{equation}
holds, where a single unit triggers the avalanche. A generic perturbation
will disturb the cluster such that it gets triggered by a single unit
only. Thus, if an avalanche of size $a$ exist, condition \eqref{eq:MFUb}
is sufficient and necessary for its stability. 

As a specific example, consider a linear partial reset function $R(\zeta)=c\zeta$.
Using equality in \eqref{eq:MFUb}, the bifurcation values $c_{\mathrm{cr}}^{(a)}$
above which a cluster state with maximal cluster size $a$ becomes
unstable are then determined implicitly by the equation \begin{equation}
e^{b\left(1-\left[\left(N-a\right)+c_{\mathrm{cr}}^{(a)}\left(a-1\right)\right]\tilde{\varepsilon}\right)}\left(e^{-b\tilde{\varepsilon}}-1\right)=\left(e^{-bc_{\mathrm{cr}}^{(a)}\tilde{\varepsilon}}-1\right).\label{eq:cImplicit}\end{equation}
Figure \ref{fig:Convex} shows an explicit example of these theoretical
predictions for the bifurcation values $c_{\mathrm{cr}}^{(a)}$ which
well match the numerical results. 

In general, we infer from \eqref{eq:cImplicit} that \begin{equation}
0<c_{\mathrm{cr}}^{(N)}<c_{\mathrm{cr}}^{(N-1)}<\dots<c_{\mathrm{cr}}^{(2)}<1\label{eq:cSequence}\end{equation}
independent of specific parameters $b$, $\tilde{\varepsilon}$ and
$N$. First, this implies that the entire sequence of bifurcations
is guaranteed to occur in the physically relevant open interval $c\in(0,1)$.
Second, with increasing $c$, states with larger clusters become unstable
before states with smaller clusters. In particular, the synchronous
state becomes unstable first and cluster states with at most two synchronized
neurons become unstable last. Third, for $a=2$ we find that the largest
bifurcation point \begin{equation}
c_{\mathrm{cr}}^{(2)}=\frac{1}{b\tilde{\varepsilon}}\ln\left(1-e^{b(1-\left(N-1\right)\tilde{\varepsilon})}\left(1-e^{b\tilde{\varepsilon}}\right)\right)\label{eq:c2}\end{equation}
can be arbitrarily small, e.g. as $b\rightarrow-\infty$. Thus the
entire sequence of desynchronizing bifurcations can occur for arbitrary
small $c$. 

\begin{figure}[t]
\begin{centering}
\includegraphics{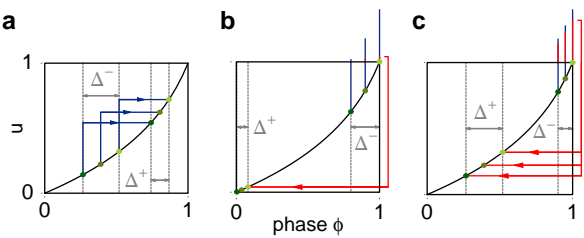}
\par\end{centering}

\caption{\label{fig:Mechanisms}Synchronizing and desynchronizing mechanisms
underlying the desynchronization transition. The phase-potential relation
are shown as solid black curves. The phase differences before ($\Delta^{-}$)
and after spike reception ($\Delta^{+}$) are indicated. \textbf{(a)}
Phase synchronization due to sub-threshold inputs coactig with either
\textbf{(b)} synchronization for strongly refractory partial resets
($c\approx0$) or \textbf{(c)} desynchronization for conservative
partial resets $(c\approx1$) determine the stability of clusters
under the return map \eqref{eq: return map}. }

\end{figure}

The from \eqref{eq: return U_b} of the return map reveals the mechanisms
underlying the desynchronization transition as the interplay between
synchronization due to sub-threshold inputs mediated by $H_{(N-a)\tilde{\varepsilon}}$
(cf. Fig. \eqref{fig:Mechanisms}a) and further synchronization or
desynchronization due to supra-threshold inputs and partial reset
mediated by $J_{a\tilde{\varepsilon}}$, depending on the strength
of the partial reset (cf. Fig. \eqref{fig:Mechanisms}b,c). The large
clusters get unstable first since they receive less synchronizing
sub-threshold inputs from the other smaller clusters and additionally
the desynchronization due to the reset is stronger in larger avalanches.

The observed desynchronization transition prevails for networks of
inhomogeneously coupled units and neurons with rise functions of mixed
convex and concave curvature, as e.g. characteristic for quadratic
integrate-and-fire neurons \cite{QIF}, the normal form of type I
excitable neurons. Moreover, our simple model system can be connected
to biophysically more detailed type I models by comparing spike time
response curves that encode the shortening of the inter-spike intervals
(ISI) following an excitatory input at different phases of the neural
oscillation. An excitatory stimulus that causes the neuron to spike
will maximally shorten the ISI in which the stimulus is applied. Additionally
the following ISI is typically affected as well. This effect can be
characterized by an appropriately chosen partial reset in our simple
system \cite{Kirst}. Networks of two-compartment conductance based
neurons indeed exhibit similar desynchronization transitions when
varying the coupling between soma and dendrite (not shown) which in
our simplified model controls the partial reset. 

In summary, we introduced a simple model of spiking neurons with \emph{partial
reset} to investigate collective network effects of possible residual
charges that may be important after somatic reset. Already for globally
and homogeneously coupled networks we find that residual charges present
after spike generation drastically affect the network dynamics. We
revealed a new desynchronization mechanism that controls a sequential
destabilization of cluster states. For no or only small fractions
$c\in\left[0,c_{\mathrm{cr}}^{(N)}\right)$ of conserved charge, the
synchronous state and cluster states with many different cluster sizes
coexist whereas for large fractions, $c\in\left(c_{\mathrm{cr}}^{(2)},1\right]$,
only the asynchronous state is left. For intermediate $c\in\left[c_{\mathrm{cr}}^{(N)},c_{\mathrm{cr}}^{(2)}\right]$
there is a sequence of bifurcations, each destabilizing the largest
stable cluster. Interestingly, this entire sequence may occur in an
interval at arbitrarily small $c>0$. 

The mechanism for neural desynchronization discussed above differs
strongly from known mechanisms that are based, e.g., on heterogeneity,
noise, or delayed feedback \cite{Desync1,DesyncDelay}. Possibly,
this novel mechanism may also be used in modified form to prevent
synchronization in neural systems like in Parkinson tremor or in epileptic
seizures \cite{DesyncDelay}. This calls for a future systematic study
of the impact of $c$ and related parameters that modulate local response
properties and thus synchronization. The simple model system introduced
above offers the first example of an analytically tractable network
model which, based on partial reset, characterizes an essential feature
of biophysically detailed compartmental models \cite{DSCoupling,BioPartialReset}.

MT acknowledges support by the Federal Ministry of Education and Research
(BMBF), Germany, under grant number 01GQ0430. CK acknowledges financial
support by the German Academic Exchange Service (DAAD).

\end{document}